\documentclass[12pt,preprint]{aastex}
\usepackage[dviprt]{}

\slugcomment{}

\shorttitle{Imaging study of HCOOCH$_{3}$ towards Orion KL}
\shortauthors{Sakai et al.}

\begin{document}


\title{ALMA imaging study 
of methyl formate (HCOOCH$_{3}$) in the torsionally excited states towards Orion KL }


\author{Yusuke Sakai, and Kaori Kobayashi,}
\affil{Department of Physics, University of Toyama,
    3190 Gofuku, Toyama, Toyama 930-8555 Japan}
\email{kaori@sci.u-toyama.ac.jp}

\and

\author{Tomoya Hirota}
\affil{National Astronomical Observatory of Japan and
 SOKENDAI (The Graduate University for Advanced
  Studies), Osawa 2-21-1, Mitaka, Tokyo 181-8588, Japan}
\email{tomoya.hirota@nao.ac.jp}




\begin{abstract}
We recently reported the first identification of rotational transitions of methyl formate (HCOOCH$_{3}$)
 in the second torsionally excited state toward Orion Kleinmann-Low (KL) observed with the Nobeyama 45 m telescope.  
 In combination with the identified transitions of methyl formate in the ground state and the first torsional excited state,
 it was found that there is a difference in rotational temperature and vibrational temperature, 
 where the latter is higher.
   In this study, high spatial resolution analysis by using Atacama Large Millimeter/Submillimeter Array
    (ALMA) science verification data was carried out to verify and understand this difference.
    Toward the Compact Ridge, two different velocity components at 7.3 and 9.1~km~s$^{-1}$ were confirmed, 
    while a single component at 7.3~km~s$^{-1}$ was identified towards the Hot Core.  The intensity maps 
    in the ground, first, and second torsional excited states have quite similar distributions.
   Using extensive ALMA data, we determined the rotational and vibrational temperatures
    for the Compact Ridge and Hot Core by the conventional rotation diagram method. 
    The rotational temperature and vibrational temperatures agree for the Hot Core and for one component of the Compact Ridge.
    At the 7.3~km~s$^{-1}$ velocity component for the Compact Ridge, the rotational temperature was found to be higher than the vibrational temperature.  
    This is different from what we obtained from the results by using the single-dish observation.
The difference might be explained by the beam dilution effect of the single-dish data
 and/or the smaller number of observed transitions within the limited range of energy levels ($\leq$30 K) of $E_u$ in the previous study.
 \end{abstract}





\section{Introduction}

Methyl formate (HCOOCH$_{3}$) was first identified toward Sgr B2 by \cite{bro75} 
and is one of the most abundant interstellar molecules that is 
almost inevitably found in star-forming regions \citep{liu01,bis07}. 
Orion Kleinmann-Low (KL) is a molecular-rich source and there are many studies on it, 
 including spectral line surveys \citep{sut85,tur91,sch97,sch01,beu05,beu06,ter10,ter11,cro14}.  
Methyl formate is one of the contributing molecules showing more than 1000 transitions \citep{lov09,tak12}.

Molecules with CH$_{3}$ rotor(s) such as HCOOCH$_{3}$ involves interaction between pure rotation and
 CH$_{3}$ internal rotation which is equivalent with torsional vibration. 
 Splittings of rotational transitions due to this internal rotation are observed \citep{gor84}. 
 The quantum number of the torsional state is represented by $v_{t}$ in this study.
 The torsional vibrations are normally low-lying vibrational modes
  so that the excited states of these modes are likely to be populated even in space
   and the pure rotational transitions in the torsional excited states may be observed.  
For methyl formate, the first and the second torsional excited state of methyl formate is located 
at about 132 cm$^{-1}$ (1 cm$^{-1}$=1.438 K; \citep{tud12}) and 250 cm$^{-1}$ \citep{sen05} above the ground state, respectively.

The laboratory microwave spectroscopy in the ground state was first reported by \cite{cur59}.  
The molecular constants, structure, and dipole moment were determined.
Many studies to extend the ground state assignments were carried out \citep{bau79,dem83,plu84,plu86,plu87,oes99,kar01}.
The torsional first excited state was studied 10 years ago \citep{oga04} and extended global analyses by other groups were also reported \citep{car07,ily09}.
\cite{mae08} identified the second torsional state as well as the ground and the first torsional excited state.  
We have also worked on the second excited state and extensively extended the assignment and analysis,
 and provided the molecular constants, energies of the state, the partition function \citep{kob13}.

Astronomically, the first excited state of methyl formate was first identified in Orion KL \citep{kob07}
 and subsequently in W 51 e2 \citep{dem08}.
 \cite{fav11} carried out a detailed study on the spatial mapping of HCOOCH$_{3}$ 
based on the observation by the Plateau de Bure Interferometer (PdBI) with the highest 
spatial resolution of 1.8" $\times$ 0.8".  
They identified 28 compact emission peaks and compared them with other spatial distributions,
 such as infrared (IR) sources, H$_{2}$ emission, and dust continuum emission.
In their study, it was pointed out that the rotation temperatures obtained from $v_{t} =$ 0 and $v_{t} =$ 1 are similar. 
 \cite{fav14} also reported the survey of HCOOCH$_{3}$ and its isotopologues H$^{13}$COOCH$_{3}$ and HCOO$^{13}$CH$_{3}$ in $v_{t} =$ 0 and 1 
  toward massive star-forming regions including Orion KL 
  where they used the science verification (SV) data of the Atacama Large Millimeter/Submillimeter Array (ALMA).
 
 We recently reported the first identification of rotational transitions of HCOOCH$_{3}$
 in the second torsionally excited state ($v_{t} = 2$; \citep{tak12}) based on the laboratory microwave spectroscopic results \citep{kob13}.
 In combination with the transitions in $v_{t} =$ 0 and 1, observed by the Nobeyama 45 m radio telescope and the IRAM 30 m telescope, 
 it was found that there is a difference in rotational temperature and vibrational temperature where the latter was higher.  
 The origin of this difference was considered to be due to a collision with hydrogen molecules and/or the far-infrared heating.  
In this paper, we present high spatial resolution ALMA SV data of methyl formate lines in Orion KL.
 We will discuss the detailed spatial distribution of methyl formate in different torsionally excited states 
 with an increased number of observed transitions in a wider range of excitation energy than those of previous single-dish results.

\section{Observations}


The public spectral line survey data, ALMA SV
 toward the Orion KL, was used in this study.  The band 6 (215-245 GHz) covers the 1 mm wavelength region. 
The tracking center position of Orion KL
was set to be R.A. = 05$^{h}$35$^{m}$14\fs35 and decl.=-05\arcdeg22\arcmin35.\arcsec 0
(J2000). The data consist of 20 spectral settings and the
net on-source time for each setting was about 20 minutes. The
baseline lengths ranged from 17 to 265 m
and consisted of 16 $\times$ 12 m antennas. The primary beam size of
each 12 m antenna is about 30\arcsec \ at band 6.
The spectral resolution of the ALMA correlator was 488 kHz,
which corresponds to the velocity resolution of 0.60-0.65 km s$^{-1}$
at the observed frequency range.  The data were analyzed by
using the Common Astronomy Software Applications (CASA)
package. The natural-weighted beam size was 1\farcs7 $\times$ 1\farcs4 with
the position angle of 171\arcdeg. 
A first order fit to the line-free channels was used to estimate the continuum emission removal. 
The Hogbom algorithm was used for the clean. 
The resultant typical rms (root-mean square)
noise level is 0.01-0.05 Jy beam$^{-1}$ for each channel map.
  



\section{Results and Discussion}
\subsection{Analysis of the spectral lines}
For the analysis of the spectra, the rest frequencies of methyl formate in $v_{t} =$0 and 1 were taken
 from the measurement at the University of Toyama.
  The estimated accuracy is 50 kHz and the assignment was done based on the JPL catalogue \citep{pik98}.
   The data set is from \cite{ily09} which includes the data by 
   \cite{bro75}, \cite{bau79}, \cite{dem83}, \cite{plu84}, \cite{plu86}, 
   \cite{oes99}, \cite{kar01}, \cite{oda03}, \cite{oga04}, \cite{car07}, and \cite{mae08}.
The rest frequencies in $v_{t} =$2 were taken from \cite{mae08} and \cite{kob13}.  
Based on these laboratory studies, many lines in ALMA SV data were identified to be due to methyl formate.
It was immediately recognized that a number of lines were contaminated 
by the other molecular transitions with a few exceptions.  
Any possible overlap of other lines was examined based on the molecular line
database, Splatalogue \citep{rem07}.  
Molecules known to exist in Orion KL were considered but were ignored 
when the expected transition intensity of the contaminated lines were weak.   
This is typical for lines with high-energy upper levels or small transition dipole moments.
  Isotopologues of known species were mostly ignored.  
Still many lines were considered to be blended.
  All identified transitions of methyl formate in $v_{t} =$0, 1 and 2 are given 
  in Tables 1 and 2 including intensities and other information.  
  Some of the lines represent the overlap of the multiple methyl formate transitions.  
   Comments are summarized as follows.
   
   B:  Blend.  Apparently, contamination by other species was acknowledged.  
   In the case of the Compact Ridge, two velocity components centered at 7.3~km~s$^{-1}$ and 9.1~km~s$^{-1}$ are identified as discussed in the next section. 
If either component is contaminated, the transition was considered to be a blended line.
   Some of the severely blended lines were excluded when extracting intensity information.
   
   NB: No blend.  When the intensity of the contaminated line is weaker than about 10\% of the methyl formate line,
    it was considered to be no blend.
   
   ND: Not detected.  
   
  Briefly, 84, 68, and 62 transitions were found to be due to the transitions of methyl formate 
  in $v_{t} =$0, 1 and 2, respectively.  
 
\subsection{Spatial Distribution and Channel Maps}   
\begin{figure}
\epsscale{0.65}
\plotone{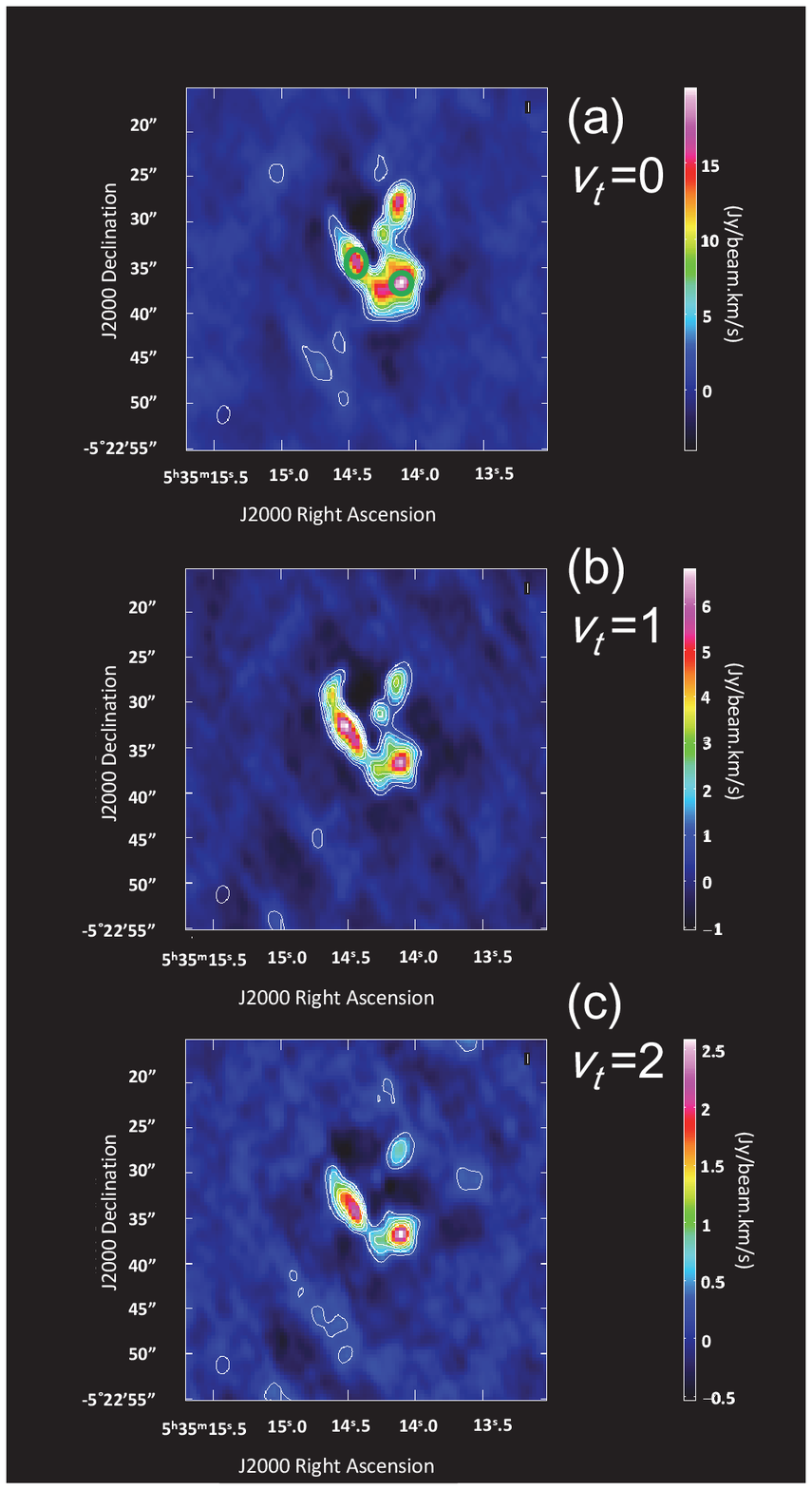}
\caption{Integrated intensity maps of methyl formate ($V_{LSR}$ 4-12 (km s$^{-1}$)).  
From top to bottom: transition of 18$_{4,15}$-17$_{4,14}$ in $v_{t}$=0, 1, and 2 at 221674 MHz (A sublevel), 219705 MHz (A sublevel) and 216594 MHz (E sublevel). 
The green circles shown in Figure 1(a) represent the Compact Ridge and Hot Core. 
The contour represents 3$\sigma$ and 
1$\sigma$s of Figures 1(a)-(c), which are 516 mJy beam$^{-1}$, 204 mJy beam$^{-1}$, and 94 mJy beam$^{-1}$, respectively. \label{imaps}}
\end{figure}

Figure \ref{imaps} shows examples of the integrated intensity maps from 4 to 12~km~s$^{-1}$ of methyl formate lines 
($J_{Ka Kc}$ = 18$_{4\ 15}$ -17 $_{4\ 14}$) in $v_{t} =$0, 1, and 2.  
We have chosen this set of transitions to compare the spatial distribution of methyl formate in different torsional states.  
For this purpose, we searched for clean transitions that have no contamination from other molecular lines. 
However, we could not find such methyl formate lines in exactly the same rotational transitions at all of the three different torsional states. 
Here, we employ the E sublevel of the $v_{t} =$2 line and the A sublevel of $v_{t} =$0 and 1 lines 
because A and E sublevels of methyl formate should have similar line intensities.
We note that the $v_{t} =$1 line is affected by a possible contamination from HC$_{3}$N. 
However, the contamination would be small because the spatial distribution is found to be similar to other clean transitions of methyl formate in the $v_{t} =$1 state. 
Although the intensity becomes weaker toward the higher vibrational states, their maps morphologically resemble those of lower vibrational states.
The spatial distributions of other methyl formate transitions also show no significant differences,
 suggesting that dependency on the $E_u$ in the range of about 130-550 K or $S\mu^{2}$ is small.
  We focused our analysis on the torsionally excited state lines
   for the Compact Ridge and Hot Core 
   where the intensity of HCOOCH$_{3}$ is strong enough to identify lines in the higher torsional excited states.
  The green circles shown in Figure 1(a) show the area of the Compact Ridge and Hot Core.  
  The center coordinates of the Compact Ridge and Hot Core are (R.A.(J2000), decl.(J2000))=
   (05h35m14s.10, -05$^{\circ}$22\arcmin36\arcsec.7) and
   (05h35m14s.45, -05$^{\circ}$22\arcmin34\arcsec.8), respectively.
  
 Figure 2 shows the channel maps of the ground state HCOOCH$_{3}$ 18$_{2,16}$-17$_{2,15}$ E line ($E_u$ =
143.23 K) at 216830.129 MHz.  
  The strong peaks have been found toward the Compact Ridge and Hot Core.  
      In the paper by \cite{fav11}, the Compact Ridge corresponds to MF1 and the Hot Core corresponds to MF2.  
      The Compact Ridge has a strong peaks in the channel maps at 7.69 and 9.04~km~s$^{-1}$ 
      while the Hot Core has a peak only at 7.69~km~s$^{-1}$.
    Basically, our spatial distribution agrees with the previous studies \citep{fav11, wid12}.  
    This spatial distribution also agrees well with the dimethyl ether distribution \citep{wid12, bro13}.
  In our data (Fig.2) we also observed two weaker peaks, that correspond to the so-called MF3 and MF4/MF5 peaks previously reported by \cite{fav11}.
  Another small peak was found between MF1 and clumps of MF4 and MF5.  
  This peak nearly corresponds to the IRc7 \citep{shu04} and was also seen by \cite{fav14}.
 
The spectra of the Hot Core can be fitted by a single Gaussian and the center of velocity is about 7.3 ~km~s$^{-1}$. 
On the other hand,
 there are two velocity components at the Compact Ridge corresponding to $V_{LSR}$
  of about 7.3~km~s$^{-1}$ and 9.1~km~s$^{-1}$ as shown in Figure 3.
     The 7.3~km~s$^{-1}$ component is mostly stronger than the 9.1~km~s$^{-1}$ component as shown in Figure 2
     but is comparable when the optically thin condition does not hold.  
     This optical depth problem will be discussed in the subsection 3.4.
These two velocity components in the Compact Ridge were also found in previous studies \citep{fav11,tak12,hir14}.  

\subsection{Partition Functions}   
   The partition function used in this study was $Q=Q_{rot}Q_{tor}$.  
   The rotational partition function was calculated 
    by using the analytical rotational partition function, which is shown below \citep{tur91}.  
\begin{equation}
Q_{rot} = 2\sqrt{\frac{\pi}{ABC}(\frac{kT}{h})^{3}}
\end{equation}
This analytical equation is appropriate when $hA<<kT$. 
Factor two represents the factor of A and E sublevels, 
   though there are small differences in the energy of these A and E sublevels.
     In this study, we have analyzed the second torsional excited state
      and modified the partition function to include the vibrational part of the first and second torsional excited state.
     The torsional excited state energy levels are considerably different from 
     what is expected from the harmonic oscillator approximation.
     Therefore, we calculated the torsional factor as follows.
     \begin{equation}
Q_{tor}=\sum^n_{v_{t}=0}\exp{-\frac{\Delta E}{kT}},
\end{equation}
where $\Delta E$ and $n$ represent the torsional energy levels and the highest torsional quantum number.  
The A sublevel rotational constants given by \cite{bau79} were used to calculate Equation (1).
As mentioned above, the energy levels of A and E sublevels are not exactly the same, 
but the difference is small compared with the torsional spacings up to $v_{t}=$2.  
Therefore, for Equation (2), $\Delta E$=132 cm$^{-1}$ \citep{tud12} and $\Delta E$=250 cm$^{-1}$ \citep{sen05}
 were assumed for $v_{t}=$1 and 2, respectively, for both the A and E sublevels. 

The partition function $Q=Q_{rot}Q_{tor}$ at different approximation levels are shown in Table 3.
The values used in this study were shown in the fourth column.
We compared the partition function provided by the JPL catalog \citep{pik98} and by \cite{fav14}. 
The JPL values were calculated by the direct sum up to the  $v_{t} =$1 
considering the $K-$level degeneracy $g_{K}$ and the reduced nuclear spin degeneracy $g_{I}$ 
which resulted in a factor of two \citep{tur91}.
This factor of two is cancelled by the choice of the intensity calculation in the rotation diagram, which is the same as that of \cite{fav14}.
  The JPL values were divided by two for direct comparison as listed in Table 3. 
The direct sum is certainly a better model, but 
the comparison at the same level shows that our adopted simple and easy-to-calculate model is within a 1$\%$ error even at 9.375 K.
   \cite{fav14} also calculated the direct sum rotational partition function and included the effect of even higher vibrational states.  
  Their calculation showed that the effect of torsional states higher than $v_{t} =$2 is less than 5$\%$ at 150 K and smaller for lower temperatures.
  Therefore, we calculated the partition function up to $v_{t} =$2 as discussed above.  
   It should also be noted that there are two other vibrational modes that could contribute the partition function other than $v_{t} =$3 and 4.
   They are the fist excited state of the COC deform mode (332 cm$^{-1}$) and the C-O torsional mode (318 cm$^{-1}$) \citep{shi72}.  
   Inclusion of these four states increases the partition function by 12$\%$ at 150 K and explains the difference by \cite{fav14} and ours.

\subsection{Rotation Diagrams}   
In order to evaluate the temperatures and column densities of HCOOCH$_{3}$ in Orion KL, the conventional rotation diagram method is employed \citep{tur91}. 
At first, the local thermodynamic equilibrium (LTE) condition and the optically thin
condition were assumed for the analysis \citep{gol78} and rotation diagrams
were prepared for Hot Core and Compact Ridge velocity components.
We found a smaller intensity for the transitions with a large $S\mu^{2}$ in $v_{t}$=0 for the 7.3~km~s$^{-1}$ component in the Compact Ridge and Hot Core.
 Thus, we concluded that it was necessary to consider the optically thick condition for these transitions.  
 It was also found that this is the case when the intensity for the 7.3~km~s$^{-1}$ component is 
 comparable to that for the 9.1~km~s$^{-1}$ component as shown in Figure 3.
We set the criteria to exclude lines with $S\mu^{2} \geq 35$ in $v_{t}$=0 for the 7.3~km~s$^{-1}$ component in the Compact Ridge and Hot Core 
to avoid the optically thick transitions. 
 Many strong, clearly detected lines in Tables 1 and 2 had to be removed to calculate temperatures and column densities for this reason.
Assuming a temperature of 150 K, which may be a little higher for the Compact Ridge but reasonable for the Hot Core,
optical depths $\tau$ of a few lines were calculated and they were about two.   
At $S\mu^{2} \simeq 35$, where the optical depth was about 2, the intensity of the 7.3~km~s$^{-1}$ component becomes small relative to that of the 9.1~km~s$^{-1}$ component.
The optically thick condition was already found in the previous survey study \citep{tur91,fav14}.  
Optically thick transitions sometimes showed skewed line profiles and the quality of a single Gaussian fit gave a somewhat poorer fit.
Therefore, optically thick transitions and contaminated transitions were removed 
   to determine the rotational temperature and vibrational temperature.  
  Removal of the optically thick transitions improved the quality of the fit.
 
 The final rotation diagrams are shown in Figure 4.
The diagram in Figure 4(d) was used to derive the effective temperature and the column density. 
This treatment implicitly assumes that the rotation and vibration are thermalized and 
hence the excitation temperature is common for all of rotational and vibrational states. 
 Our effective temperature calculated by using all of the data is equivalent to the vibrational temperature by \cite{tak12}.
The rotational temperature in each vibrational state and the vibrational temperature for each component are compiled in Table 4
  with previous results \citep{fav11,tak12,fav14}.  
At the Compact Ridge, the effective temperature of the two components are similar and roughly 90 K.
The temperature is higher (about 120 K) for the Hot Core.  
It was not possible to determine the realistic rotational temperature of HCOOCH$_{3}$ in $v_{t} =$2 for the Hot Core.
This is due to the fact that numerous lines are blended toward the Hot core as opposed to the Compact Ridge.  
However, the data of HCOOCH$_{3}$ in $v_{t}=$2 for the Hot Core were used to determine the effective temperature.
The rotational temperature is higher than the vibrational temperature for the Compact Ridge 7.3~km~s$^{-1}$ component. 
The rotational temperature and vibrational temperature for the Hot Core and Compact Ridge 9.1~km~s$^{-1}$ component match within its errors. 
The column density for the Compact Ridge 7.3~km~s$^{-1}$ component is three times larger than 
that for the Compact Ridge 9.1~km~s$^{-1}$ component.
 Figure 4(d) shows all the data in one panel and clearly shows the tendency. 

In the previous study \citep{tak12}, the vibrational temperature was clearly higher than the rotational temperature.    
Note that the observation by \cite{tak12} was carried out with the Nobeyama 45 m single-dish telescope and the area averaged temperature was obtained 
where the Compact Ridge and Hot Core were not spatially resolved.
 The difference may be attributed to the beam dilution. In addition, 
  we employed many transitions with wider range ($\geq$100 K) of $E_u$ compared with the previous study \citep{tak12}.
 It is likely that the relatively narrower range ($\leq$30 K) of $E_u$ employed in the previous study \citep{tak12} 
 may result in larger uncertainties to derive excitation temperatures.  
 To evaluate the column density, \cite{tak12} considered the partition function including the second torsional excited state.  
 Still it was smaller than ours by one-two orders of magnitude.
 This fact may also be explained by the effect of the beam dilution.

\cite{fav11} also determined the rotational temperature where lines of multi-torsional states were used
 so that it is the same as our effective temperature.  
  They used a very limited number of transitions to drive temperatures and column densities 
  so that the errors are sometimes very large or sometimes underestimated.  
  Considering this fact, their values are thought to agree with ours.
The column densities by \cite{fav11} are smaller than ours. 
  As was pointed out by \cite{fav14}, this may be due to the fact 
  that up to the first torsional state was considered in the partition function used by \cite{fav11} 
  and the new results \citep{fav14} are similar to our results (50\% difference only).

\section{Summary}

The ground, first, and second torsional excited states of HCOOCH$_{3}$ in Orion KL 
were identified using ALMA SV data, and images of spatial distribution were examined. 
The distribution of these states are quite similar and all of the states have bright peaks toward the Compact Ridge and Hot Core.
  Toward the Compact Ridge, two velocity components were recognized, which were found previously \citep{fav11,tak12,hir14}.
The vibrational (torsional) temperature and the rotational temperature for the Hot Core and Compact Ridge
 were determined by using the conventional rotation diagram method. 
The vibrational temperature and rotational temperature for the Hot Core and Compact Ridge 9.1~km~s$^{-1}$ component 
agreed within their errors, contrary to the previous study \citep{tak12}
 where the vibrational temperature was higher than the rotational temperature.  
 In addition, the rotational temperature is higher than the vibrational temperature for the Compact Ridge 7.3~km~s$^{-1}$ component.  
The difference may be explained by the beam dilution and/or the fact that 
 small number of transitions with narrow ranges ($\leq$30 K) of $E_u$ were considered in the previous study \citep{tak12}.



\acknowledgments

This paper makes use of the following ALMA data: ADS/JAO.ALMA\#2011.0.00009.SV. 
ALMA is a partnership of ESO (representing its member states), NSF (USA) and NINS (Japan),
 together with NRC (Canada) and NSC and ASIAA (Taiwan), in cooperation with the Republic of Chile. 
The Joint ALMA Observatory is operated by ESO, AUI/NRAO, and NAOJ. 
This study was partly supported by a Grant-in-Aid for Scientific Research on Innovative Areas
 by the Ministry of Education, Culture, Sports, Science, and Technology of Japan (grant numbers 24684011, 25108005, and 26108507). \\

\begin{rotate}
\begin{deluxetable}{lccccccccccccccccccccc}
\tabletypesize{\tiny}
\tablewidth{0pt}
\tablecaption{Transition List for the Compact Ridge
\label{tab-lines}}
\tablehead{
\colhead{}&   \colhead{}&   \colhead{}&   \colhead{} &   \colhead{}&   \colhead{}
&   \colhead{}&   \colhead{}&   \colhead{}&   \colhead{}&   \colhead{}
&\multicolumn{4}{c}{7.3~km~s$^{-1}$ Component} &   \colhead{} &   \multicolumn{4}{c}{9.1~km~s$^{-1}$ Component}&   \colhead{}&   \colhead{}    \\
\cline{12-15} \cline{17-20} \\
\colhead{Frequency}       & \colhead{$v_{t}$ }       & \colhead{$J'$} & \colhead{$Ka'$}& \colhead{$Kc'$}
& \colhead{$J"$} & \colhead{$Ka"$}& \colhead{$Kc"$} & \colhead{Symmetry} & \colhead{$S\mu^{2}$} &
\colhead{$E_u$} &\colhead{$T$}&\colhead{FWHM}&\colhead{$V_{LSR}$}&\colhead{$W$}
&   \colhead{}&\colhead{$T$}&\colhead{FWHM}&\colhead{$V_{LSR}$}&\colhead{$W$}&\colhead{Comments}&\colhead{Diagram}\\
\colhead{(MHz)}       & \colhead{}       & \colhead{} & \colhead{}& \colhead{}
& \colhead{} & \colhead{}& \colhead{} & \colhead{} & \colhead{($D^{2}$)} &
\colhead{(K)} &\colhead{(K)}&\colhead{(km s$^{-1}$)}&\colhead{(km s$^{-1}$)}&\colhead{(K km s$^{-1}$)}
&   \colhead{}&\colhead{(K)}&\colhead{(km s$^{-1}$)}&\colhead{(km s$^{-1}$)}&\colhead{(K km s$^{-1}$)}&\colhead{}&\colhead{}
}
\startdata
215972.058 	&	0	&	19	&	1	&	18	&	18	&	2	&	17	&	A	&	7.9 	&	145.2 	&	38.34	&	1.57 	&	7.35 	&	64.17 	&		&	12.11	&	1.59 	&	9.12 	&	20.52 	&	NB	&	*	\\
216210.859 	&	0	&	19	&	1	&	18	&	18	&	1	&	17	&	E	&	47.5 	&	145.2 	&	71.31	&	1.78 	&	7.16 	&	135.06 	&		&	60.04	&	2.23 	&	8.92 	&	142.34 	&	NB	&		\\
216830.129 	&	0	&	18	&	2	&	16	&	17	&	2	&	15	&	E	&	44.4 	&	143.2 	&	72.13	&	1.73 	&	7.13 	&	133.06 	&		&	60.46	&	2.18 	&	8.91 	&	140.06 	&	NB	&		\\
216964.812 	&	0	&	20	&	1	&	20	&	19	&	1	&	19	&	E	&	50.8 	&	147.2 	&		&		&		&		&		&		&		&		&		&	B	&		\\
217215.780 	&	0	&	32	&	9	&	24	&	32	&	8	&	25	&	A	&	10.6 	&	325.2 	&	9.89	&	1.83 	&	7.06 	&	19.25 	&		&	4.57	&	1.96 	&	9.30 	&	9.53 	&	NB	&		\\
218280.833 	&	0	&	17	&	3	&	14	&	16	&	3	&	13	&	E	&	41.9 	&	139.3 	&	94.70	&	1.94 	&	7.29 	&	195.60 	&		&	63.19	&	1.72 	&	9.11 	&	115.63 	&	NB	&		\\
218297.853 	&	0	&	17	&	3	&	14	&	16	&	3	&	13	&	A	&	41.9 	&	139.3 	&	84.19	&	1.73 	&	7.17 	&	155.36 	&		&	66.80	&	2.12 	&	8.96 	&	150.71 	&	NB	&		\\
218585.436 	&	0	&	36	&	9	&	28	&	36	&	8	&	29	&	A	&	12.2 	&	383.3 	&	2.95	&	1.48 	&	7.26 	&	4.66 	&		&	2.01	&	2.88 	&	8.48 	&	6.15 	&	B	&		\\
\enddata
\tablenotetext{a}{Abbreviations in the comment column are B:blend; NB: not blended; and ND: no lines found. 
An asterisk indicates that the optically thin line was used in the rotation diagram analysis.}
\end{deluxetable}

\begin{deluxetable}{lcccccccccccccccc}
\tabletypesize{\scriptsize}
\tablewidth{0pt}
\tablecaption{Transition List for the Hot Core
\label{tab-lines}}
\tablehead{\colhead{Frequency}       & \colhead{$v_{t}$ }       & \colhead{$J'$} & \colhead{$Ka'$}& \colhead{$Kc'$}
& \colhead{$J"$} & \colhead{$Ka"$}& \colhead{$Kc"$} & \colhead{Symmetry} & \colhead{$S\mu^{2}$} &
\colhead{$E_u$} &\colhead{$T$}&\colhead{FWHM}&\colhead{$V_{LSR}$}
&\colhead{$W$}&\colhead{Comments}&\colhead{Diagram}\\
\colhead{(MHz)}       & \colhead{}       & \colhead{} & \colhead{}& \colhead{}
& \colhead{} & \colhead{}& \colhead{} & \colhead{} & \colhead{$D^{2}$} &
\colhead{(K)} &\colhead{(K)}&\colhead{(km s$^{-1}$)}&\colhead{(km s$^{-1}$)}
&\colhead{(K km s$^{-1}$)}&\colhead{}&\colhead{}}
\startdata
215972.058 	&	0	&	19	&	1	&	18	&	18	&	2	&	17	&	A	&	7.9 	&	145.2 	&	11.07 	&	1.88 	&	7.54 	&	22.15 	&	NB	&		\\
216210.859 	&	0	&	19	&	1	&	18	&	18	&	1	&	17	&	E	&	47.5 	&	145.2 	&	47.53 	&	2.99 	&	7.05 	&	151.07 	&	NB	&		\\
216830.129 	&	0	&	18	&	2	&	16	&	17	&	2	&	15	&	E	&	44.4 	&	143.2 	&	48.01 	&	3.09 	&	6.96 	&	157.83 	&	NB	&		\\
216964.812 	&	0	&	20	&	1	&	20	&	19	&	1	&	19	&	E	&	50.8 	&	147.2 	&		&		&		&		&	B	&		\\
217215.780 	&	0	&	32	&	9	&	24	&	32	&	8	&	25	&	A	&	10.6 	&	325.2 	&	5.16 	&	2.27 	&	7.17 	&	12.48 	&	B	&		\\
218280.833 	&	0	&	17	&	3	&	14	&	16	&	3	&	13	&	E	&	41.9 	&	139.3 	&	51.96 	&	3.23 	&	7.04 	&	178.88 	&	NB	&		\\
218297.853 	&	0	&	17	&	3	&	14	&	16	&	3	&	13	&	A	&	41.9 	&	139.3 	&	51.98 	&	3.08 	&	6.99 	&	170.44 	&	NB	&		\\
218585.436 	&	0	&	36	&	9	&	28	&	36	&	8	&	29	&	A	&	12.2 	&	383.3 	&		&		&		&		&	B	&		\\
\enddata
\tablenotetext{a}{Abbreviations in the comment column are B:blend; NB: not blended; and ND: no lines found. 
An asterisk indicates that the optically thin line was used in the rotation diagram analysis.}
\end{deluxetable}
\end{rotate}

\begin{deluxetable}{lrrrrr}
\tabletypesize{\scriptsize}
\tablewidth{0pt}
\tablecaption{Partition function for methyl formate at different approximations\label{tbl-3}}
\tablehead{
\colhead{Temperature (K)}       & \colhead{$Q$ $v_{t}$=0} &   \colhead{$Q$ $v_{t}$=0, and 1} &
\colhead{$Q$ $v_{t}$=0, 1, and 2} & \colhead{\shortstack{JPL Catalog\\ $Q$ $v_{t}$=0, and 1 \tablenotemark{a}}} 
 & \colhead{\shortstack{\cite{fav14}\\ $Q$ $v_{t}$=0 -6 and Other Vibrational Levels}}
 }
\startdata
300	&	64720.28	&	99096.14	&	118622.03	&	99801.35& 249172.44	\\
225	&	42037.06	&	60119.20	&	68625.21	&	60551.01& 104015.96	\\
150	&	22882.07	&	29337.45	&	31420.20	&	29536.48& 36433.43	\\
75	&	8090.04	&	8733.91	&	8800.94	&	8774.41& 8894.06	\\
37.5	&	2860.26	&	2878.38	&	2878.57	&	2886.21& 2885.30	\\
18.75	&	1011.25	&	1011.29	&	1011.29	&	1015.42& 1015.31	\\
9.375	&	357.53	&	357.53	&	357.53	&	360.41&	360.33\\
\enddata
\tablenotetext{a}{The JPL catalog values include nuclear spin factor of 2.  For direct comparison, these values shown here were calculated from the JPL values divided by 2. }
\end{deluxetable}

\begin{rotate}
\begin{deluxetable}{lrrrrrrrr}
\tabletypesize{\scriptsize}
\tablewidth{0pt}
\tablecaption{Comparison of temperatures and column densities of methyl formate \label{tbl-4}}
\tablehead{
\colhead{Position}       & \colhead{\shortstack{ $T_{\rm rot}$ (K) \\$v_{t}$=0} }       & \colhead{\shortstack{ $T_{\rm rot}$ (K)\\$v_{t}$=1}} &
 \colhead{\shortstack{$T_{\rm rot}$ (K)\\$v_{t}$=2}} & \colhead{\shortstack{$T_{\rm eff}$ (K)\\ \space}} & \colhead{\shortstack{$N_{\rm tot}$\\ (cm$^{-2}$)}} &
\colhead{\shortstack{$T_{\rm eff}$(K) \tablenotemark{a}\\ \space}} 
& \colhead{\shortstack{$N_{\rm tot}$\\ (cm$^{-2}$) \tablenotemark{a}}}& \colhead{\shortstack{$N_{\rm tot}$\\ (cm$^{-2}$) \tablenotemark{b}}}
}
\startdata
\shortstack{Compact Ridge \\7.3~km~s$^{-1}$ component}  & 146$\pm$17 & 153$\pm$20& 219$\pm$56&  99.6$\pm$8.1 &1.11$\pm$0.31$\times10^{18}$& 79$\pm$2 &1.60$\pm$0.10$\times10^{17}$&5-8.5$\times10^{17}$ \\
\shortstack{Compact Ridge \\9.1~km~s$^{-1}$ component}  & 82.9$\pm$3.7  & 93$\pm$14& 97$\pm$28     &  86.8$\pm$4.1 &3.42$\pm$0.79$\times10^{17}$& 112$\pm$50 &6.8$\pm$2.1$\times10^{16}$&7$\times10^{16}$ \\
Hot Core&  162$\pm$26     & 153$\pm$24 & -         & 119.9$\pm$9.2 &6.0$\pm$1.6$\times10^{17}$& 128$\pm$9  &1.60$\pm$0.20$\times10^{17}$& 3.3-4.3$\times10^{17}$ \\
Averaged \tablenotemark{c}&  43$\pm$9   & 53$\pm$8 & -          & 124$\pm$5 &1.38$\pm$0.14$\times10^{16}$ &-         &-                      &-\\
\enddata
\tablecomments{In the column head, $T_{\rm rot}$, $T_{\rm eff}$, and $N_{\rm tot}$ represent rotational temperature, effective temperature, and column density.
  Effective temperature is defined the temperature by using data in multiple torsional states. }
\tablenotetext{a}{\cite{fav11}}.
\tablenotetext{b}{\cite{fav14}}.
\tablenotetext{c}{\cite{tak12}}.
\end{deluxetable}
\end{rotate}

\clearpage



\clearpage

\begin{figure}
\epsscale{1.1}
\plotone{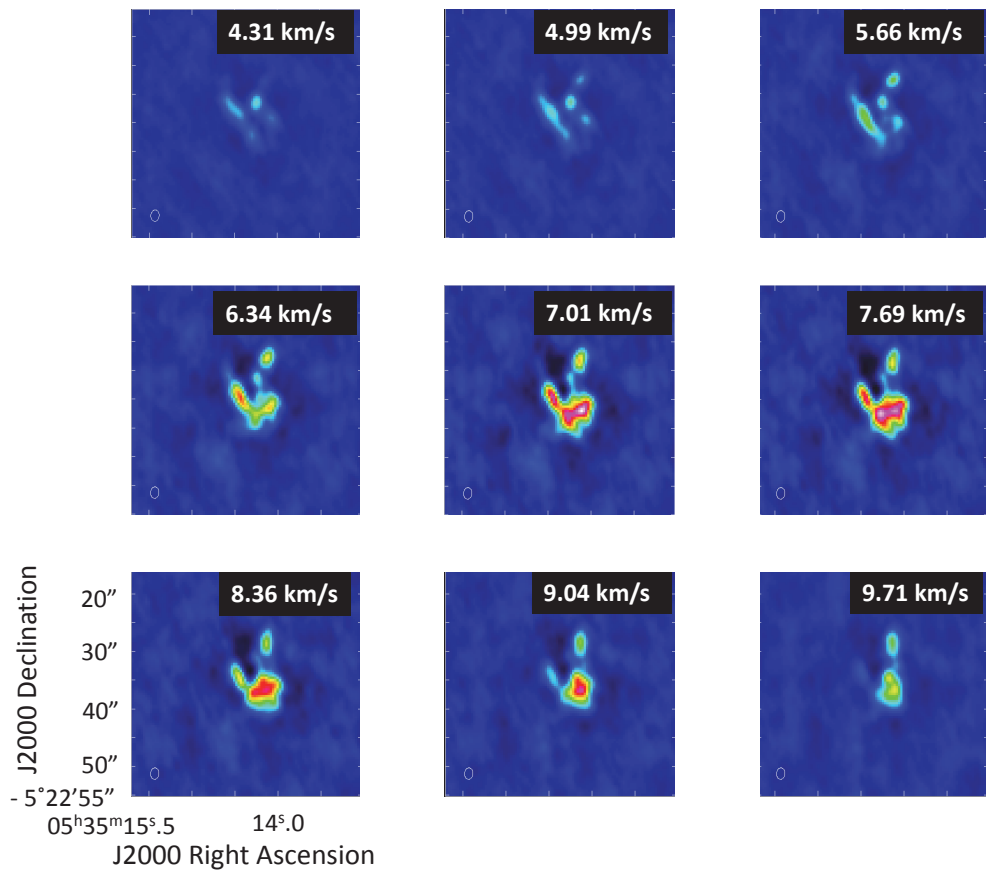}
\caption{Channel maps of the ground state methyl formate 18$_{2,16}$-17$_{2,15}$ E line in $v_{t}$=0 ($E_u$(K) =
143.23 K) at 216830.129 MHz}
\end{figure}

\clearpage


\begin{figure}
\plotone{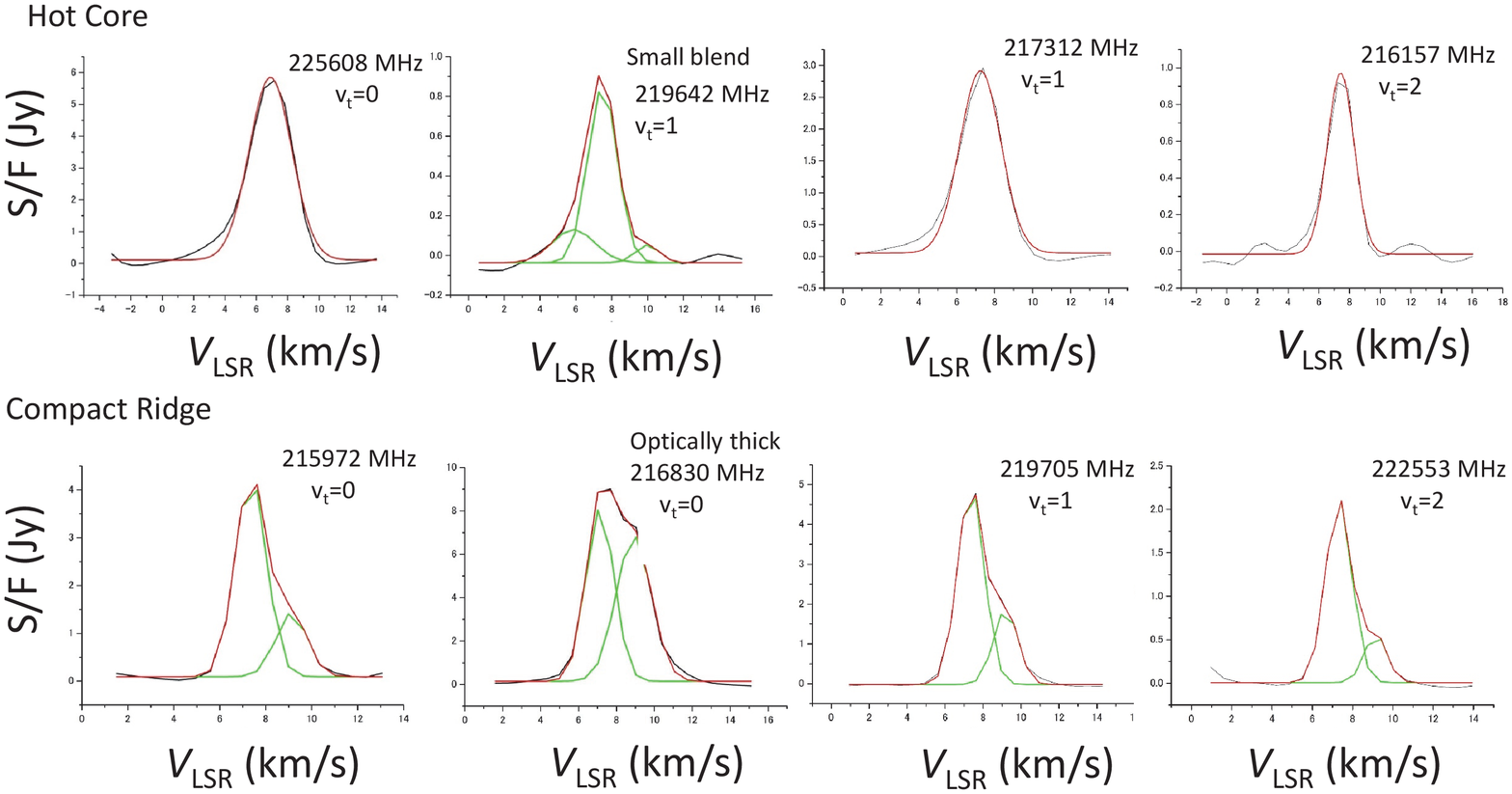}
\caption{Examples of spectra. 
Green, red, and black lines represent the each fitted velocity component, total, and observation.  
We have also modeled the contaminant in the second panel in the top left.
}
\end{figure}

\begin{figure}
\plotone{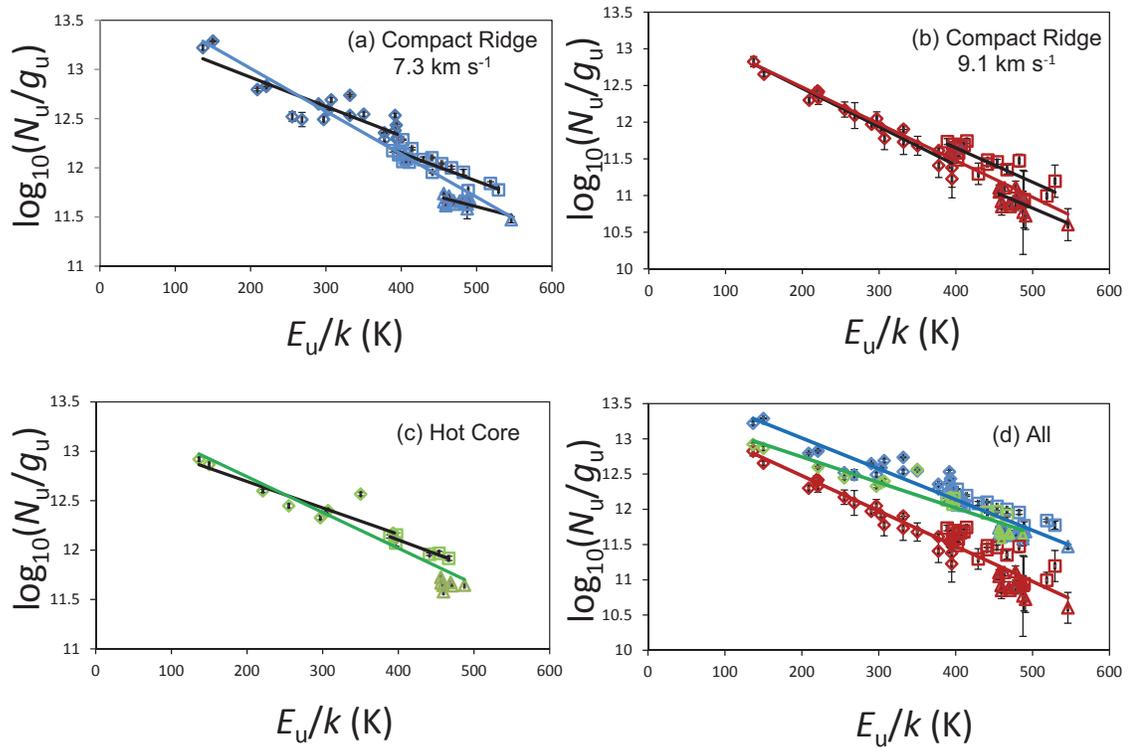}
\caption{Rotation diagram of methyl formate at the two positions.  
There are two velocity components at the Compact Ridge.  
Blue, red, and green colors
 represent transitions for the Compact Ridge 7.3~km~s$^{-1}$ component, the 9.1~km~s$^{-1}$ component, and the Hot Core. 
The diamonds, rectangles, and triangles represent $v_{t}$=0, 1, and 2, respectively.  
The corresponding colored lines show the least-square fitted lines to derive the effective temperatures by using all the transitions.  
Black colored lines were used to derive the rotational temperature of the each torsional states.
The error bars were calculated from the errors of intensity of the Gaussian fit.  }
\end{figure}






\clearpage


\begin{thebibliography}{}
\bibitem[Bauder (1979)]{bau79} 	Bauder, A. 1979, J. Phys.  Chem. Ref. Data, 8, 583
\bibitem[Beuther et al.(2005)]{beu05} Beuther, H., Zhang, Q., Greenhill, L. J., et al, 2005, \apj, 632, 355
\bibitem[Beuther et al.(2006)]{beu06} Beuther, H., Zhang, Q., Reid, M. J., et al, 2006, \apj, 636, 323
\bibitem[Bisschop et al.(2007)]{bis07} Bisschop, S. E., J\o rgensen, J. K., van Dishoeck, E. F., \& de Wachter, E. B. M. 2007, \aap, 465, 913
\bibitem[Brouillet et al.(2013)]{bro13} Brouillet, N., Despois, D., Baudry, A., et al., 2013, \aap, 550, A46
\bibitem[Brown et al.(1975)]{bro75} Brown, R. D., Crofts, J. G., Godfrey, P. D., et al, 1975, \apjl, 197, L29
\bibitem[Carvajal et al.(2007)]{car07} Carvajal, M., Willaert, F., Demaison, J., \& Kleiner, I. 2007, J. Mol. Spectrosc., 246, 158
\bibitem[Crockett et al.(2014)]{cro14} Crockett, N. R., Bergin, E. A., Neill, J. L., et al,
    2014, \apj, 787, 112
\bibitem[Curl (1959)]{cur59} 	Curl, R. F., Jr. 1959, J. Chem. Phys, 30, 1529
\bibitem[Demaison et al.(1983)]{dem83} Demaison, J., Boucher, D., Dubrulle, A., \& Van Eijck, B. P. 1983, J. Mol. Spectrosc., 102, 260
\bibitem[Demyk et al.(2008)]{dem08} Demyk, K., Wlodarczak, G., \& Carvajal, M. 2008, A\&A, 489, 589
\bibitem[Favre et al.(2011)]{fav11} Favre, C., Despois, D., Brouillet, N., al., 2011, \aap, 532, A32
\bibitem[Favre et al.(2014)]{fav14} Favre, C., Carvajal, M., Field, D., et al., 2014, \apjs, 215, 25
\bibitem[Hirota et al.(2014)]{hir14} Hirota, T., Tsuboi, M., Kurono, Y., et al., 2014, \pasj, 66, 106
\bibitem[Ilyushin et al.(2009)]{ily09} 	Ilyushin, V.; Kryvda, A.; \& Alekseev, E. 2009, J. Mol. Spectrosc., 255, 32
\bibitem[Lovas(2009)]{lov09} Lovas, F. J., NIST Recommended Rest Frequencies for Observed Interstellar
 Molecular Microwave Transitions 2009 Revision, Gaithersberg:NIST, http://physics.nist.gov/PhysRefData/Micro/Html/contents.html
\bibitem[Goldsmith \& Langer(1978)]{gol78} Goldsmith, P. F., \& Langer, W. D., 1978, \apj, 222, 881
\bibitem[Gordy \& Cook(1984)]{gor84}  Gordy, W., \& Cook, R. L. 1984, Microwave Molecular Spectra 3rd ed., New
York: Wiley
\bibitem[Karakawa et al.(2001)]{kar01} Karakawa, Y., Oka, K., Odashima, H., Takagi, K., \& Tsunekawa, S. 2001, J. Mol.
Spectrosc., 210, 196
\bibitem[Kobayashi et al.(2007)]{kob07} Kobayashi, K., Ogata, K., Tsunekawa, S., \& Takano. S.
    2007, \apjl, 657, L17
\bibitem[Kobayashi et al.(2013)]{kob13} Kobayashi, K., Takamura, K., Sakai, Y., et al,    2013, \apjs, 205, 9
\bibitem[Liu et al.(2001)]{liu01} 	Liu, S.-Y., Mehringer, D. M., \& Snyder, Lewis E. 2001, \apj, 552, 654
\bibitem[Maeda et al.(2008)]{mae08} Maeda, A., De Lucia, F.C., \& Herbst, E. 2008, J. Mol. Spectrosc., 251, 293
\bibitem[Odashima et al.(2003)]{oda03} Odashima, H., Ogata, K., Takagi, K.,
    \& Tsunekawa, S.  2003, Molecules, 8, 139
\bibitem[Oesterling et al.(1999)]{oes99} Oesterling, Lee C., Albert, S., De Lucia, F. C., Sastry, K. V. L. N. \& Herbst, E. 1999, \apj, 521, 255
\bibitem[Ogata et al.(2004)]{oga04} Ogata, K., Odashima, H., Takagi, K., \& Tsunekawa, S. 2004, J. Mol.
Spectrosc., 225, 14
\bibitem[Pickett et al.(1998)]{pik98}Pickett, H. M., Poynter, R. L., Cohen, E. A., Delitsky, M.L., Pearson, J. C., \& M\"{u}ller, H.S.P.
 1998, J. of Quant. Spectrosc. Rad. Trans., 60, 883
\bibitem[Plummer et al.(1984)]{plu84} Plummer, G. M., Herbst, E., De Lucia, F.,  \& Blake, G. A. 1984, \apjs, 55, 633
\bibitem[Plummer et al.(1986)]{plu86} Plummer, G. M., Herbst, E., De Lucia, F. C.,  \& Blake, G. A. 1986, \apjs, 60, 949
\bibitem[Plummer et al.(1987)]{plu87} Plummer, G. M., Herbst, E., \& De Lucia, F. C. 1987, \apj, 318, 873
\bibitem[Remijan et al.(2007)]{rem07}Remijan, A. J., Markwick-Kemper, A., \& ALMA Working Group on Spectral Line Frequencies, \baas, 39, 963
\bibitem[Schilke et al.(1997)]{sch97}Schilke, P., Groesbeck, T. D., Blake, G. A., \& Phillips; T. G.
1997, \apjs, 108, 301
\bibitem[Schilke et al.(2001)]{sch01}Schilke, P., Benford, D. J., Hunter, T. R., Lis, D. C., \& Phillips; T. G.
2001, \apjs, 132, 301
\bibitem[Senent et al.(2005)]{sen05}Senent, M. L., Villa, M., Mel\'endez, F. J., \& Dom\'inguez-G\'omez, R.
2005, \apj, 627, 567
\bibitem[Shimanouchi (1972)]{shi72}Shimanouchi, T. 1972, Tables of Molecular Vibrational Frequencies Consolidated Volume I, 1
\bibitem[Shuping et al.(2004)]{shu04}Shuping, R. Y., Morris, M., \& Bally, J. 2004, \aj, 128, 363
\bibitem[Sutton et al.(1985)]{sut85}Sutton, E. C., Blake, G. A., Masson, C. R., \& Phillips, T. G. 1985, \apjs, 58, 341
\bibitem[Takano et al.(2012)]{tak12} Takano, S., Sakai, Y., Kakimoto, S., Sasaki, M., \& Kobayashi, K. 2012, \pasj, 64, 89
\bibitem[Tercero et al.(2010)]{ter10} Tercero, B., Cernicharo, J., Pardo, J. R., \& Goicoechea, J. R. 2010, \aap, 517, A96
\bibitem[Tercero et al.(2011)]{ter11} Tercero, B., Vincent, L., Cernicharo, J., Viti, S.; \& Marcelino, N. 2011, \aap, 528, A26
\bibitem[Tudorie et al.(2012)]{tud12}Tudorie, M., Ilyushin, V., Vander Auwera, J., Pirali, O., Roy, P., \& Huet, T. R.
2012, JCP, 137, 064304
\bibitem[Turner(1991)]{tur91} Turner, B. E., 1991, \apjs, 76, 617
\bibitem[Widicus Weaver \& Friedel(2012)]{wid12} Widicus Weaver, S. L. \& Friedel, D. N. 2012, \apjs, 201, 16
\end{thebibliography}
\end{document}